\documentclass[twocolumn,superscriptaddress,aps,prl,showpacs,10pt]{revtex4-1}
\usepackage{amsmath}
\usepackage{siunitx}
\usepackage{amssymb}
\usepackage{color}
\usepackage{graphicx}
\usepackage{bm}
\usepackage{dcolumn}
\usepackage{xcolor}
\usepackage{mhchem}
\usepackage[caption=false]{subfig}
\newcommand{\mpsr}{$\mu^+$SR}
\newcommand{\cvo}{Co$_2$VO$_4$}
\usepackage[T1]{fontenc}
\usepackage{tabularx,booktabs,multirow,array}
\begin{document}




\title{Localized-delocalized crossover of spin-carriers and magnetization reversal in Co$_{2}$VO$_{4}$}

\author{Abhijit Bhat Kademane}
\affiliation{Department of Mathematics and Physics, Universitetet i Stavanger, 4036 Stavanger, Norway}
\author{Churna Bhandari}
\affiliation{Ames Laboratory, Iowa State University, Ames, Iowa 50011, USA}
\altaffiliation[A.B.K and C.B\,]
{contributed equally.}
\author{Durga Paudyal}
\affiliation{Ames Laboratory, Iowa State University, Ames, Iowa 50011, USA}
\affiliation{Department of Electrical and Computer Engineering, Iowa State University, Ames, Iowa 50011, USA}
\author{Stephen Cottrell}
\affiliation{ISIS Facility, STFC Rutherford Appleton Laboratory, Harwell Science and Innovation Campus, Chilton, Didcot, Oxon OX11 0QX, UK}
\author{Pinaki Das}
\author{Yong Liu}
\author{Yuen Yiu}
\affiliation{Ames Laboratory, Iowa State University, Ames, Iowa 50011, USA}
\author{C. M. Naveen Kumar}
\affiliation{Institute of Solid State Physics, Vienna University of Technology, Wiedner Hauptstrae 810, 1040 Vienna, Austria}
\affiliation{AGH University of Science and Technology, Faculty of Physics and Applied Computer Science, 30-059 Kraków, Poland}
\author{Konrad Siemensmeyer}
\author{Andreas Hoser}
\affiliation{Helmholtz Zentrum Berlin f\"ur Materialien und Energie, D-14109 Berlin, Germany}
\author{Diana Lucia Quintero-Castro}
\affiliation{Department of Mathematics and Physics, Universitetet i Stavanger, 4036 Stavanger, Norway}
\author{David Vaknin}
\affiliation{Ames Laboratory, Iowa State University, Ames, Iowa 50011, USA}
\affiliation{Department of Physics and Astronomy, Iowa State University, Ames, Iowa 50011, USA}
\author{Rasmus Toft-Petersen}
\affiliation{Department of Physics, Technical University of Denmark, DK-2800 Kongens Lyngby, Denmark}

\date{\today}

\begin{abstract}

Neutron diffraction, magnetization and muon spin relaxation measurements, supplemented by density functional theory (DFT) calculations are employed to unravel temperature-driven magnetization reversal (MR) in inverse spinel Co$_2$VO$_4$. All measurements show a second order magnetic phase transition at $T_{\rm C} = 168$\,K to a  collinear ferrimagnetic phase. The DFT results suggest the moments in the ferrimagnetic phase are delocalized and undergo gradual localization as the temperature is lowered below $T_{\rm C}$. The delocalized-localized crossover gives rise to a maximum magnetization at $T_{\rm NC} = 138$\,K and the continuous  decrease in magnetization produces sign-change at  $T_{\rm MR} \sim 65$\,K. Muon spectroscopy results support the DFT, as  a strong $T_1$-relaxation is observed around $T_{\rm NC}$, indicating highly delocalized spin-carriers gradually tend to localization upon cooling. The  magnetization reversal determined at zero field is found to be highly sensitive to applied magnetic field, such that above $B\sim 0.25$\,T instead of a reversal a broad minimum in the magnetization is apparent at $T_{\rm MR}$.  Analysis of the neutron diffraction measurements shows two antiparallel  magnetic sub-lattice-structure, each belonging to magnetic ions on two distinct crystal lattice sites. The relative balance of these two structural components in essence determines the magnetization. Indeed, the order parameter of the magnetic phase on one site develops moderately more than that on the other site. Unusual tipping of the magnetic balance, caused by such site-specific magnetic fluctuation, gives rise to a spontaneous flipping of the magnetization as the temperature is lowered. 

\end{abstract}

\maketitle

\section{Introduction}
The $AB_2X_4$-type spinel compounds, with tetrahedral $A$-site and octahedral $B$-site, have been an anchoring point for some of the earliest work on magnetism that led to the seminal N\'eel \cite{Neel1948} and the Yafet-Kittel models \cite{Yafet1952}.  The spinel (and the inverse-spinel) compounds display a complex interplay among lattice, spin, charge, and orbital degrees of freedom \cite{TSURKAN2021}, where novel phenomena emerge \cite{Yamasaki2006,Ruff2019}. Indeed, interest in these systems  persists as they exhibit exotic magnetic phenomena, such as multiferroic effects and the spin liquid state that result from spin-orbit coupling and geometric frustration associated with the pyrochlore sub-lattice of the $B$-site \cite{Balents2010,Bergman2007}. 
A prototypical class of spinel compounds is the vanadate of type $A$V$_2$O$_4$ and its inverse-spinel counterpart $A_2$VO$_4$. At high temperatures, these systems form a cubic structure with space-group $Fd\bar{3}m$. For $A$V$_2$O$_4$ with a non-magnetic transition metal ion, such as Mg$^{2+}$ or Zn$^{2+}$ in the $A$-site, a  cubic-to-tetragonal phase transition accompanied by an antiferromagnetic phase have been reported \cite{Niitaka2013,Lee2004}. On the other hand, magnetic ions on the $A$-site, such as in \ce{MnV2O4}, \ce{FeV2O4}, \ce{CoV2O4}, show a multitude of structural and magnetic transitions by virtue of coupled spin- and orbital- degrees of freedom \cite{Garlea2008,MacDougall2012,Zhang2016,Ishibashi2017,Nii2012}. Specifically, for \ce{CoV2O4}, orbital glass-like behavior has been reported \cite{Koborinai2016,Plessis2016} and {\it spin-ice rules} have been outlined in studies of \ce{MnV2O4} and \ce{FeV2O4}  \cite{Garlea2008,MacDougall2012}. The inverse-spinel vanadate counterparts (of type $A_2$VO$_4$), which are more complex due to random-cation distribution,   have yet to be thoroughly explored. Indeed, the intriguing magnetization reversal (MR) \cite{KUMAR2015} has been observed in inverse spinel Co$_2$VO$_4$ \cite{Menyuk1960}. It has been proposed that the cation distribution of Co$_2$VO$_4$ is Co$_A^{2+}$(Co$^{2+}_\nu$V$^{4+}_{1-\nu})_B$O$_4^{2-}$ \cite{Rogers1963}. Additionally, magnetic interaction mechanisms in Co$_2$VO$_4$ have been proposed using semi-empirical rules for magnetic interactions in spinels \cite{Menyuk1960,Wickham1959}. While temperature-driven magnetization reversal for \ce{Co2VO4}, \ce{Co2TiO4}, and \ce{Sr2DyRuO6} has been experimentally explored \cite{Menyuk1960,Sakamoto1962,Thota2017,Nayak2015,Adroja2020}, the fundamental driving mechanism behind MR in the inverse spinels remains to be understood. 

Here, we examine the temperature evolution of magnetic structures using neutron diffraction, magnetization, polarized muon spectroscopy, and electronic structure calculations of \ce{Co2VO4}. We establish multiple phase transitions in the magnetically ordered phase, including temperature-driven MR. We find the structure to be composed of two nearly balanced antiparallel ferromagnetic components, on the $A$- and $B$-site, giving rise to a net ferrimagnetic phase evolving with temperature. Performing density functional theory (DFT), {\it localized} calculation representing low-temperatures and {\it delocalized} calculations for high-temperature (within the ordered region), we confirm that the MR in Co$_2$VO$_4$ is consistent with such a balance. Polarized muon spectroscopy results show dynamic $T_1$-relaxation at high-temperature (within the ordered region) supporting electron delocalization predicted by the theory.

\section{Experimental and Computational Details}
The polycrystalline sample of \cvo \ was prepared using a solid-state reaction as described in Ref. \citep{Becker1956}. The X-ray diffraction (XRD) measurement confirms the formation of \ce{Co2VO4}\, spinel. Rietveld refinement of neutron diffraction and XRD data (at 300 K) confirm the cubic $Fd\bar{3}m$ symmetry. Due to $A/B$ site inversion ($A$ -- $8b$-tetrahedral and $B$ -- $16c$-octahedral) the formal chemical formula of Co$_2$VO$_4$  is Co$^{2+}_\nu$ V$^{4+}_{1-\nu}$[V$^{4+}_\nu$ Co$^{2+}_{2-\nu}$]O$_4$, where $\nu$ is the inversion parameter, and the $B$-site cations are enclosed in brackets. The refinement yields that the degree of inversion $\nu\simeq1$, i.e., the actual Co$^{2+}_A$[V$^{4+}$Co$^{2+}$]$_B$O$_4$, where Co$^{2+}$ occupies the $A$ - site and the remaining Co$^{2+}$ and V$^{4+}$ are distributed evenly in the  $B$ site.   Unlike the common spinel counterparts \cite{Garlea2008,Ishibashi2017}, we find no structural transitions in \cvo\ at all measured temperatures. Nonetheless, we observed an insignificant shrinking of the unit-cell that appears to saturate at $\approx$ 65\,K.

\begin{figure}
\includegraphics[width=0.5\textwidth]{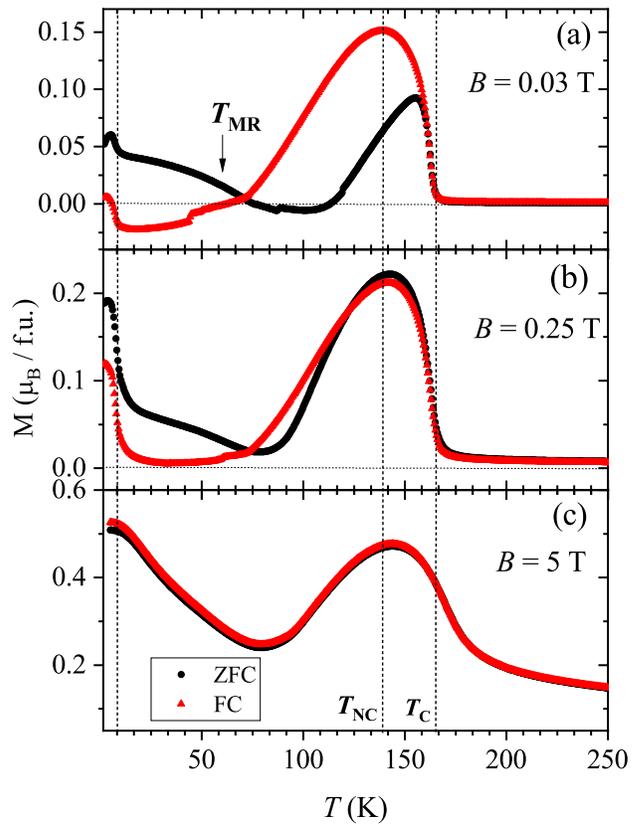}
\caption{\label{magnetization}Temperature dependence of ZFC (black circle) and FC (red triangle) magnetization measurements performed for applied magnetic fields 0.03\,T, 0.25\,T and 5\,T. $T_{\rm {C}}$ and $T_{\rm {NC}}$ represent the transition temperatures to collinear and non-collinear magnetic structures respectively, as explained in text. Magnetization reversal temperature is represented by $T_{\rm {MR}}$ for $B= 0.03$\,T.}
\end{figure}

\begin{figure*} 
\includegraphics[width=1\textwidth]{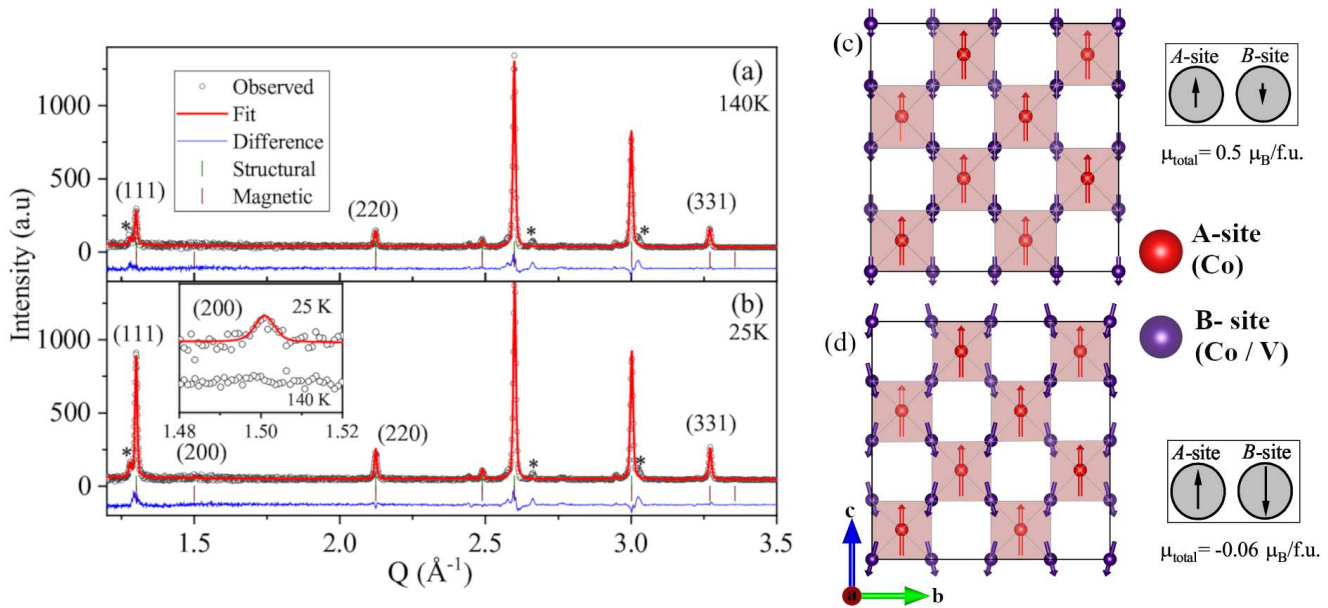}
\caption{Neutron-diffraction patterns (black open circles), fits from the Rietveld refinement (red solid lines), and their differences (blue solid lines). The vertical bars are the expected Bragg peak positions as mentioned in the panels. The asterisk indicate impurity peaks. NPD observed (measured on POWGEN) (a) $T$ = 140~K and (b) $T$ = 25~K corresponding to the data collected with center wavelengths 1.333\ \AA.  In the inset of (b) one can see the emergence of (200) between 140\,K and 25\,K. (c) Magnetic structure of Co$_{2}$VO$_{4}$ at 140\,K. (d) Magnetic structure at 25\,K. $A$-site is occupied by Co(red), whereas $B$-site occupancy is shared between Co and V(purple).}
\label{diffractionM} 
\end{figure*} 

\begin{figure}
\includegraphics[width=0.51\textwidth]{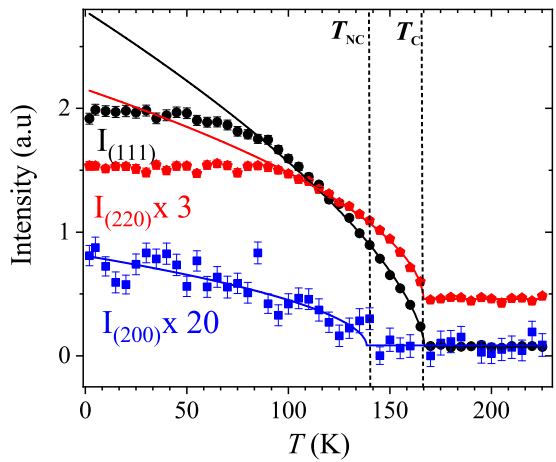}
\caption{\label{orderparameter1}Temperature dependencies of integrated intensities of the Bragg peaks (111),(220), and (200) reflections for $B=0$\,T. The solid lines represents respective Power-law fits. $T_{\rm {C}}$ features can be identified from the ordering of (111) and (220) while $T_{\rm {NC}}$ from ordering of (200).}
\end{figure}

Magnetization measurements were performed on a pellet of mass 6.15\,mg and carried out using a superconducting quantum interference device (Quantum Design MPMS, SQUID) magnetometer for various applied magnetic fields in the range of $0.03-7$\,T and $2-350$\,K, performing both zero-field cooled (ZFC) and field cooled (FC) measurements. Neutron powder diffraction (NPD) was performed on a 2\,g polycrystalline sample on the time-of-flight (TOF) powder diffractometer, POWGEN \cite{Huq2011}, located at the Spallation Neutron Source at Oak Ridge National Laboratory. The measurements were carried out with neutrons of central wavelengths 1.333 and 3.731\,\AA. The sample was loaded in a vanadium can which was attached to the cold end of a cryostat to reach temperatures in the range $2 \le T \le 300$~K. Neutron powder diffraction in applied magnetic fields was performed using E6 \cite{Buchsteiner2009} and E9 \cite{Franz2017} diffractometers at the BER II reactor at the Helmholtz-Zentrum Berlin. The 3.5\,g powder sample was dispersed in deuterated methanol-ethanol mixture, which freezes into an amorphous glass upon cooling, and thereby fixes the powder orientation while applying a magnetic field. The E6 diffractometer was used to perform the temperature scans ranging from 15-200\,K  with an applied magnetic field of 0.25\,T and 2\,T with wavelength 2.44\,\AA. Refinements of the diffraction data were carried out using \texttt{FullProf} \cite{FullProf,Rodriguez-Carvajal2001}. 

The EMU spectrometer \cite{Giblin2014} at the ISIS Neutron and muon source was used to perform zero applied field muon spin relaxation ( ZF -\mpsr)  and longitudinal field muon spin relaxation (LF-\mpsr) measurements where the applied magnetic field and muon spin are collinear on the polycrystalline sample. The polycrystalline Co$_{2}$VO$_{4}$  \ sample amounting to 2.5\,g was placed in a titanium holder with 24\,mm diameter window. Muons falling outside the sample window were stopped in a silver mask, giving a non-relaxing background. 100\% spin-polarized muons are produced by the source, implanted in the sample and their time-evolution measured. The time evolution of muon polarization is given by $P_z(t)$. A normalized sample polarization immediately after muon implantation, $P_s(0)$, was obtained after background correction. The ZF-\mpsr \ scans were performed at various temperatures in the 10-250\,K range. In-order to elucidate dynamic aspects, LF-\mpsr \ measurements were performed at 250, 125, 75, and 10\,K with applied fields of 0.005, 0.05, 0.2, and 0.3\,T. The data were analyzed using the muon analysis module of \texttt{Mantid} \cite{Mantid}. 

To understand the magnetization reversal in Co$_2$VO$_4$, we carried out two sets of first-principles electronic structure calculations: i) {\it delocalized} and ii) {\it localized}. The delocalized calculations were performed using DFT in conjunction with generalized gradient approximation (GGA) for exchange-correlation functional in all-electron projected augmented plane wave (PAW) formalism implemented in Vienna Simulation Package (\texttt{VASP}) including spin-orbit coupling (SOC) \cite{VASP1,VASP2}. We used plane wave cut-off energy of 500 eV and $4\times4\times 4$ {\bf k}-mesh for the Brillouin-zone integration. Structure optimization is performed by randomly replacing 50\% \ce{V_B} with \ce{Co_B} in the bulk CoV$_2$O$_4$ structure that has the lowest total energy among all the considered configurations. 

\section{Results and Discussions}

Figure~\ref{magnetization} shows the temperature ($T$) dependence of the zero field-cooled (ZFC) and field-cooled (FC) magnetization ($M$) of \ce{Co2VO4} at applied magnetic field $B$ = 0.03, 0.25, and 5\,T. Three features are identified for $B = 0.03$\,T. First, a sharp rise is observed at $T_{\rm {C}}=168(1)$\,K (C-collinear) corresponding to ferrimagnetic response and second, with an onset of a decrease  at $T_{\rm {NC}}=138(8)$\,K (NC- non collinear), for both FC and ZFC. The third feature occurs at the temperature range $T_{\rm {MR}}\sim 65 - 130$\,K,  where the magnetization reverses, dependent on the cooling protocol (i.e., FC or ZFC). Also, below 20\,K there is an upturn in magnetization, suggestive of a meta-magnetic  behavior. Magnetization reversal does not occur for fields higher than $\sim 0.25$\,T and eventually, the ZFC and FC curves overlap for fields higher than 5\,T and the temperature dependence becomes reversible. 

\begin{figure}
\includegraphics[width=0.5\textwidth]{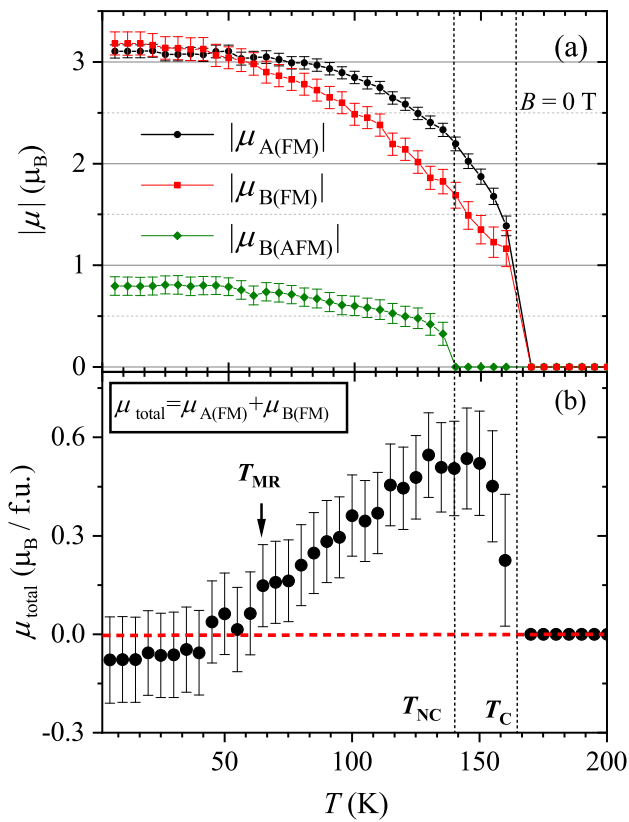}
\caption{\label{orderparameterZF}(a) The magnitude of $A$- and $B$- sub-lattice moments obtained from Rietveld refinement for $B=0$\,T. (b)Total moment (per f.u.) calculated by adding moments FM moments. Temperature dependence of total moment emulates the magnetization reversal as observed in magnetization measurement.}
\end{figure}

The magnetization at $T_{\rm {C}}$ for $B= 0.1$\,T is $\SI{0.2}{\micro_{\rm{B}}}/f.u.$ (not shown), which is lower than that of \ce{CoV2O4} \cite{Huang_2012}. A high spin configuration $S=3/2\,(\SI{3}{\micro_{\rm{B}}})$ for \ce{Co_A^2+} and  $S = 1$ $(\SI{2}{\micro_{\rm{B}}})$ for \ce{V_B^3+} have been reported for \ce{CoV2O4} \cite{Ishibashi2017}. This indicates the reduced magnetization in \ce{Co2VO4} is due to antiparallel \ce{Co_A} and \ce{Co_B}/\ce{V_B} (ferri-magnetic) assuming high spin configuration $S = 3/2$, for Co$^{2+}$ and $S = 1/2$ for \ce{V_B} implying an oxidation state of $4+$ for V. As  discussed below, DFT calculations confirm the valence states of \ce{Co_B} and \ce{V_B} as well their magnetic ordering in the {\it localized solution}. 

The neutron powder diffraction patterns for 140\ K and 25\ K are presented in Fig.\ \ref{diffractionM} (a) and (b). Consistent with the  aforementioned magnetic susceptibility, Fig.\ \ref{orderparameter1} shows the emergence of the (111) and (220)  magnetic reflections at $T=168(1)$\ K.  At a lower temperature ($=138(2)$\ K), a weak (200) magnetic reflections is also observed (inset of Fig.\ \ref{diffractionM} (a)). We emphasize that the (200) is forbidden by $Fd\bar{3}m$. In the case of \ce{MnV2O4} observation of the (200) is linked to spin canting of the spinel-$B$-site and a structural transition \citep{Garlea2008}. The temperature dependence of the three magnetic Bragg reflections is shown in Fig.\ \ref{orderparameter1}. These curves are fitted to a power-law function. The critical temperature obtained for reflections (111) and (220) is $T_{\rm {C}}$=167.8$\pm$0.4\,K and for the (200) at $T_{\rm {NC}}$ = 138.5$\pm$7.5\,K. We also note that the critical exponents associated with (111) and (200) are different, at 0.71$\pm$0.02 and 0.5$\pm$0.1, respectively.

 \begin{figure}
\includegraphics[width=0.5\textwidth]{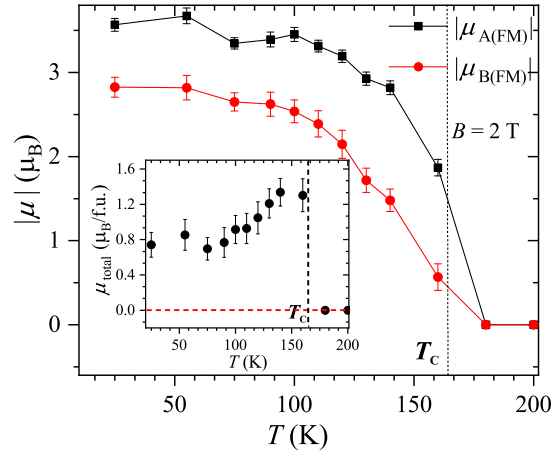}
\caption{\label{orderparameterB}(a) Magnitude of $A$- and $B$- sub-lattice moments obtained from Rietveld refinement for $B=2$\,T. (Inset) Total moment (per f.u.) calculated by adding moments along the $c$-direction. Temperature dependence of total moment emulates magnetization for fields higher than 0.25\,T.}
\end{figure}

\begin{table}[!htbp]
\setlength{\tabcolsep}{1pt}
\renewcommand{\arraystretch}{1.1}
\caption{Magnetic structure factors computed are presented as $|$F$_{hkl}|^2$ (normalized) for the magnetic reflections ($hkl$) using Basis vectors (BVs) of the Irreducible representation(IR) belonging o $A$-site and $B$-site.} 
\centering 
\begin{tabular}{c|c| c| c | c | c | c | c  } 
\hline\hline 
 Site& Irrep &\multicolumn{2}{c|}{BVs} &$|$F$_{111}|^2$ &$|$F$_{200}|^2$ & $|$F$_{220}|^2$  & $|$F$_{331}|^2$ 
\\ 
\hline 
$A$-site& $\Gamma 10$ &$\psi_A^1$    & (0 0 1)&& & &  \\ 
8b &&(FM)& (0 0 1)& 0.68 &	0 &	1.00 &	0.27 	 \\ \hline
 &     &    & (0 0 1)&  & & & \\ 
 & & $\psi_B^1$   & (0 0 1)&  & & &   \\ 
& &(FM)& (0 0 1)&0.51 &	0 & 0 &	0.2 \\ 
$B$-site& $\Gamma 10$  &         & (0 0 1)&  & & &  \\ \cline{3-8}
16c& &         & (-1 -1 0)&  & & &    \\ 
&&$\psi_B^2$        & (1 1 0)&  & & &   \\ 
 & &(AFM)& (-1 1 0)& 0.1 &	0.04&	0.03&	0.04 \\ 
 & &        & (1 -1 0)&  & & &   \\ 
 
\hline \hline 
\end{tabular}
\label{tab:StructF}
\end{table} 

To restrict the refinement of the neutron powder data, we exclude the irreducible basis vectors, on both the $A$-and $B$-site, that are inconsistent with the strongest observed peaks in the powder diffraction pattern. These basis vectors that are (for $Fd\bar{3}m$ and propagation vector $k = (0,0,0)$)) founds using \texttt{BasIREPS} \cite{FullProf} and \texttt{SARAh}\cite{Sarah}. We find that the simplest model consistent with both magnetization and powder diffraction data, consists of basis vectors from the IR $\Gamma 10$ with $A$- and $B$- sub-lattice moments arranged anti-parallel. The magnetic representation in this case corresponds to Shubnikov space group $I4_1/am’d’$ (representation mGM4+) \cite{Perez-Mato2015,Stokes}. This picture is consistent with observations of other spinel vanadates \cite{Garlea2008,Ishibashi2017} as well as recommendations made for magnetic interactions in \ce{Co2VO4} \cite{Menyuk1960}.  Table\ \ref{tab:StructF} lists  $|$F$_{hkl}|^2$ for prominent magnetic reflections calculated using these basis vectors.  Note that the intensity on (200) could be explained only by $\psi_B^2$ (AFM), while (220) has no contributions from $\psi_B^1$ (FM). As seen in table\ \ref{tab:StructF}, ferromagnetic order on the $A$-site ($\psi_A^1$) gives rise to intensity in both the (220) and (111) reflections while ferromagnetic order on the $B$-site ($\psi_B^1$) only gives rise to intensity on (111). The ferromagnetic ordered moment on the $A$-site can therefore be explicitly distinguished from that of the $B$-site, and the relative magnitudes of the ordered moments can be directly extracted from the refinement. While the net ferrimagnetic moment can be directly deduced from neutron powder diffraction, it cannot distinguish individual contributions  from \ce{Co_B} and \ce{V_B}.  We note that combination of  $\psi_B^1$ (c-axis) and $\psi_B^2$ ( ab-pane) leads to the commonly observed 2-in-2-out configuration in pyrochlore structures \cite{Garlea2008}.

The refinements of neutron powder diffraction data at ($B=0$ T) at the temperature range $5-250$ K yield the sub-lattice magnetic moments as shown in Fig.\ \ref{orderparameterZF} (a). Refined moments on $A$- and $B$-sites are labeled as $\mu _{ \mathrm{A}(\mathrm{FM})}$, $\mu _{ \mathrm{B}(\mathrm{FM})}$ and $\mu _{\mathrm{B}(\mathrm{AFM})}$, respectively. The total moment $\mu _{\rm {total}}= \mu _{ \mathrm{A}(\mathrm{FM})} + \mu _{ \mathrm{B}(\mathrm{FM})}$ is shown in Fig.\ \ref{orderparameterZF} (b). Contributions from $\mu _{\mathrm{B}(\mathrm{AFM})}$ ( AFM see Tab.\ \ref{tab:StructF}) cancel out and thus do not contribute to the total moment. As shown, $\mu_{\rm {total}}$ as a function of temperature exhibits magnetic behavior that resembles the magnetization shown in Fig.\ \ref{magnetization} (a). Indeed Fig.\ \ref{orderparameterZF} (b) shows three identified temperatures in magnetization, $T_{\rm {C}}$, $T_{\rm {NC}}$, and $T_{\rm {MR}}$. Similar analysis of the diffraction patterns under magnetic field $B=2$\,T also yield distinguishable $\mu _{ \mathrm{A}(\mathrm{FM})}$ and $\mu _{ \mathrm{B}(\mathrm{FM})}$. However,  $\mu_{\rm {total}}$ at $B=2$\,T does not show magnetization reversal (Fig.\ \ref{orderparameterB}) as $\mu _{ \mathrm{A}(\mathrm{FM})}$ is always larger than $\mu _{ \mathrm{B}(\mathrm{FM})}$ for all temperatures, consistent with the observation above $0.25$\,T (Fig.\ \ref{magnetization}). It is not possible to extract $\mu _{\mathrm{B}(\mathrm{AFM})}$ reliably in these measurements, as the defused signal from ethanol-methanol mixture overshadows the weak (200) peak.

\begin{figure}
\includegraphics[width=0.5\textwidth]{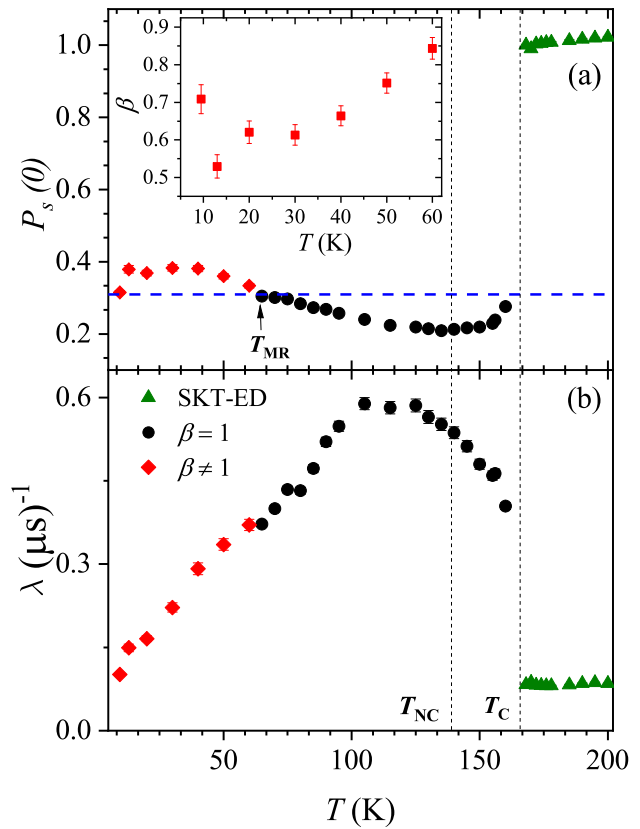}
\caption{\label{muonfits}Initial polarization (a) and relaxation rate (b) extracted from fits to the ZF-\mpsr\ polarization spectra. For $T <T_{\rm {C}}$, fits are to Eq. (\ref{Eq:muon}), with $\beta$ as a free parameter below $T_{\rm {MR}}$ (red points). Above $T_{\rm {C}}$, fits are carried out to Eq. (\ref{Eq:muon1})\,(green points). Inset of (a) $\beta$ vs $T$ where $\beta$ reaches 0.5 at 12\,K.}
\end{figure}
 
To determine possible dynamic origin of the distinct critical behavior on the $A$- and $B$-sites, employ polarized muon spectroscopy. The ZF-\mpsr\  measurements were performed at various points in the region 10$-250$\,K. In the paramagnetic region ($T>T_{\rm {C}}$), the ZF-\mpsr \ polarization spectra shows temperature-independent relaxation with full polarization. This region is parameterized with static Kubo–Toyabe times exponential decay (SKT-ED), accounting for nuclear and electronic moments \cite{Kubo1967}.

\begin{multline}
P_z(t) =  P_{z_1}(0)\left(e^{-\delta^{2} t^{2} / 2}\left(1-\delta^{2} t^{2}\right) \frac{2}{3}+\frac{1}{3}\right) e^{-\lambda t} \\
+ P_{z_2}(0), 
\label{Eq:muon1}
\end{multline}

where $\delta$ accounts for relaxation rate due to the coupling of nuclear moments and muon and $\lambda$ for relaxation rate due to electron moments and muon coupling.
At $T_{\rm {C}}$, a sharp loss of the initial polarization is observed. This is typical for a magnetically ordered system, where the missing polarization is associated with strong electronic moments, giving rise to an unresolved precession signal owing to the finite muon pulse width at the source (ISIS). Below $T_{\rm {C}}$, a one-component stretched exponential decay (Eqn. \ref{Eq:muon}) function was sufficient to parameterize the ordered region, i.e.,

\begin{equation}
    P_z(t) =  P_{z_1}(0)e^{-(\lambda t)^\beta} + P_{z_2}(0). 
\label{Eq:muon}
\end{equation}

In Eq. (\ref{Eq:muon}), $\lambda$ is the relaxation rate and $\beta$- the stretching exponent. When $\beta=1$, equation \ref{Eq:muon} turns into simple exponential decay. We find that above $T_{\rm {MR}}$, the best fit to the data is obtained with $\beta$ fixed at 1. However, below $T_{\rm {MR}}$, acceptable fits could only be obtained fitting $\beta$ as a free parameter

Figure \ref{muonfits} shows the initial polarization and relaxation rate extracted from ZF-\mpsr\ data as a function of temperature. The polarization  parameter ($P_s(0)$) decreases to 2/3 of its initial value below $T_{\rm {C}}$. Intriguingly, there is a further loss of polarization as the temperature is reduced below $T_{\rm {C}}$, reaching a minimum at $T_{\rm {NC}}$ and only regaining the 1/3 value of the initial polarization at $T_{\rm {MR}}$. The broad dip between $T_{\rm {C}}$  and $T_{\rm {MR}}$ suggests that muons are experiencing changes in the magnetic structure right through the region. This is consistent with the corresponding peak in the relaxation rate, suggesting spin fluctuation/reorientation process is driving an enhanced $T_1$-relaxation of the muon signal through this temperature region. Notably, there is no sharp peak in the relaxation rate at $T_{\rm {C}}$, as one would expect for a typical magnetic phase transition. Fitting the low-temperature data (T$<60$ K) with $\beta$ as a free parameter shows $\beta$ decreasing steadily, reaching a value $\approx 0.5$ at 12\,K. We note that $\beta = 0.5$ is a signature of glassy transition \cite{Campbell1994}. The glassy transition at low temperatures is typical of inverse spinels \cite{Nayak2015,Thota2013}. 

\begin{figure}
\includegraphics[width=0.5\textwidth]{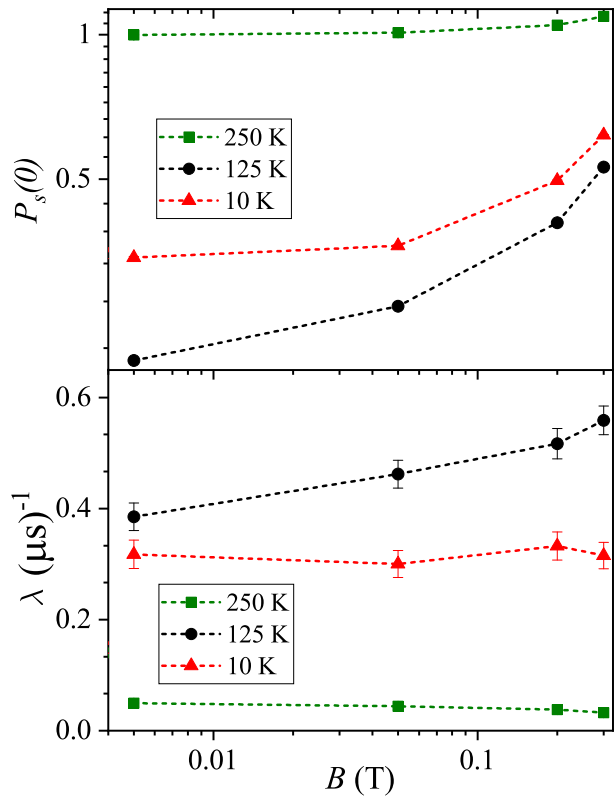}
\caption{\label{muonLFs}(a) Initial polarization (b) relaxation rate extracted from fits to the LF-\mpsr\ polarization spectra with Eq. (\ref{Eq:muon}). At 125\,K, complete decoupling was realized with tiny field of 0.005\,T. Due to instrument effects $P_s(0)$ goes beyond 1 for higher fields.}
\end{figure}

Initial polarization and relaxation data for LF-\mpsr\ measurements up to 0.3\,T, carried out at 10\,K, 125\,K and 250\,K, are shown in Figure \ref{muonLFs}. With $T_1$-spin relaxation, $\lambda$, persisting at all temperatures to the highest fields measured. This suggests that spin dynamics are present at all temperatures, both in the paramagnetic regime above $T_{\rm {C}}$ and also in the ordered state. We note that the strongest relaxation is measured at 125\,K, close to $T_{\rm {MR}}$, a region where electron delocalization is predicted by DFT, and at this temperature the relaxation rate (Fig. \ref{muonLFs}b) appears to increase with the applied field in contrast to results at measured at 10\,K. A clear recovery of the initial polarization is seen at the highest fields measured, although full decoupling is not achieved, with only ~50\% of the polarization being recovered. We suggest this is the result of both large internal fields and spin dynamics below $T_{\rm {C}}$ in this material.

To supplement the experimental results, we conduct {\it delocalized} and {\it localized} electronic structure calculations. For the delocalized calculations, we employ standard DFT with GGA functional. 
In the localized calculations, we add the on-site electron correlation following the Dudarev {\it et al.}'s scheme \cite{Dudarev} in which only the effective value of on-site correlation, ($U_{eff}=U-J = 4.5$ eV for $3d$ electrons of Co and V \cite{HubbardU}) is meaningful. To be on equal footing, we also performed another calculation using the same value of $U_{eff}$ and find the delocalized solution, which has higher total energy than in the localized solution. As expected, the net magnetic moments do not change significantly with the use of $U_{eff}$ in the delocalized results.

As shown below, the {\it localized} calculations are akin to the low-temperature ground state whereas the {\it delocalized} corresponds to the high-temperature ferrimagnetic phase.  Both {\it localized} and {\it delocalized} calculations show similar magnetic moments for \ce{Co_A}, while they are drastically different for \ce{Co_B} and \ce{V_B}. These changes in the local magnetic moments are consistent with the low-temperature neutron diffraction results (Table \ref{tab:Magn}). The major contribution to the net magnetic moment is from spin (Table~\ref{tab:Magn}), while the orbital contributions are small and slight fluctuations for \ce{Co_B} are found (not shown). In the localized solution, we find canted spin-structures for \ce{Co_B} and \ce{V_B} in qualitative agreement with the experimental results at 25 K (Table \ref{tab:Magn}). A notable prediction is the antiferromagnetic alignment of \ce{Co_A} moments with respect to \ce{Co_B} with a small canting in the planar direction (Table \ref{tab:Magn}). For \ce{Co_A}, the canting is not resolved experimentally. At the $B$-site, the \ce{V_B} moments follow the \ce{Co_B} moments. These are remarkable changes in the frustrated magnetic structures of the well-known spinel vanadates viz., FeV$_2$O$_4$ \cite{MacDougall2012}  and MnV$_2$O$_4$ \cite{Garlea2008}, in which \ce{V_B} moments are 2-in 2-out as observed experimentally. The calculated net magnetic moment parallel to the $c$-direction is negative, in agreement with the low-temperature data.  
The discrepancy between theory and experiment in the magnitude of net magnetic moment is likely due to the choice of atomic sphere sizes used in the calculations, i.e., \ce{Co_A} moment $\sim \SI{2.6}{\micro_{\rm{B}}}$ is smaller than the experiment $\SI{3.07}{\micro_{\rm{B}}}$. We find $3d$ Co/V and O($2p$) hybridization, which we argue may affect the magnetic moments on both Co and V.  
In calculations, the canting is small which results in a large magnetic moment ($\sim \SI{3.7} {\micro_{\rm{B}}}$) along the $z$-direction, which qualitatively agrees with the experiment (see Table \ref{tab:Magn}). Another possibility for the slight discrepancy is random occupancy of Co and V in the $B$-sites.

Interestingly, the magnetic moments obtained with the delocalized solutions agree with the higher temperature data, indicating a more itinerant scenario \cite{BhandariPRM021}. The magnetic moments on the $B$-site are sensitive to the choice of $U_{eff}$ (i.e., localized or delocalized), while the moments on the $A$-site remain relatively unchanged. 
Inter-atomic charge transfer from \ce{Co_B} to \ce{V_B} sites reduces the effective \ce{Co_B} moments. This delicate balance between \ce{Co_A}, \ce{Co_B}, and \ce{V_B} magnetic moments in the localized and delocalized limits presumably leads to the MR.

\begin{figure}
\includegraphics[scale=0.45]{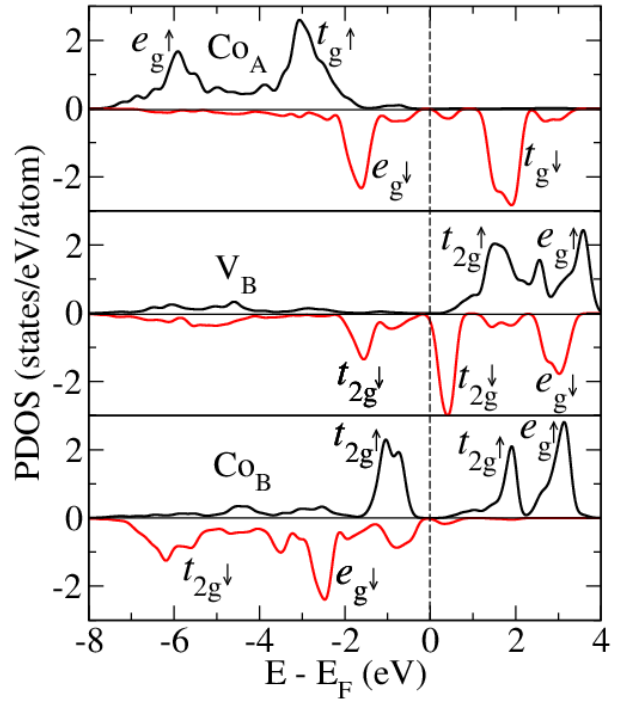}
\caption{\label{dos}Spin resolved partial density of states (PDOS) of \ce{Co_A}, \ce{Co_B}, and \ce{V_B}. The positive and negative values indicate PDOS of majority and minority spin electrons, respectively. The spins of \ce{Co_A} are aligned antiferromagnetically with respect to \ce{Co_B} and \ce{V_B}. Both Co at $A$- and $B$ sites have $3d^7$ valence-electrons indicating the formation of Co$^{2+}$ ion. The distorted octahedral crystal field splits $t_{2g}$ into an $a_{1g}$ and doubly degenerate $e_g$ states at site $B$. For \ce{Co_B}, two electrons occupy the lower $e_g$ states forming $S=3/2$ state, while for \ce{V_B}, one electron occupies $a_{1g}$ yielding V$^{4+}$ oxidation state with $S=1/2$.}
\end{figure}

To demonstrate that the electronic configuration of \ce{Co} is $3d^7$ (i.e., Co$^{2+}$) and that of \ce{V_B} is $3d^1$ (i.e., V$^{4+}$), we examine the partial density of states (PDOS) of each atom in the localized calculations. 
Figure \ref{dos} shows very well split-off $e_g\downarrow$ and $t_g\uparrow$ $3d$-states of \ce{Co_A} around the Fermi level due to the tetrahedral crystal field of oxygen atoms. In \ce{Co_B} and \ce{V_B}, the splittings of the three-fold  $t_{2g}$-states are less obvious. Because of the distorted octahedral crystal field effect\cite{KismarahardjaPRL011}, these states split into an $a_{1g}$ and doubly degenerate $e_g^{'}$ resulting in $3d^7$, $S=3/2$ (occupying $e_g^{'} \uparrow$) for \ce{Co_B}$^{2+}$ as in \ce{Co_A}$^{2+}$ and $3d^1$, $S=3/2$ (occupying $a_{1g}$) for \ce{V_B}$^{4+}$ in the localized limit
as inferred from the current and the earlier experiments \cite{Menyuk1960}. An interesting change is in vanadium oxidation state from V$^{3+} (3d^2)$, which is common in the standard spinel compounds, to V$^{4+} (3d^1)$ with $S=1/2$ consistent with the magnetic moment \cite{Menyuk1960}. 
We note that \ce{Co_B}$^{2+}$ and \ce{V_B}$^{2+}$ moments change significantly in the delocalized solutions (Table \ref{tab:Magn}). This is expected due to a narrow $t_{2g}$-splitting in both atoms at $B$-sites.

 \begin{table*}[!htbp]
\renewcommand{\arraystretch}{1.3}
\caption{Canted magnetic moment components (spin + orbital) ($\mu_x,\mu_y,\mu_z$) of individual atom and total magnetic moments ($\mu_{\rm {total}}$) (per f.u.) in units of $\SI{}{\micro_{\rm{B}}}$ in Co$_2$VO$_4$ obtained from Rietveld refinement from zero field data and theory (GGA + SOC and GGA + SOC + U calculations). For oxygens, only the calculated moments are given.} 
\begin{ruledtabular}
\begin{tabular}{c l c   c  c  c } 
 Site &Atom  & \multicolumn{2}{c}{Experiment} &\multicolumn{2}{c}{Theory}  \\
\cline{3-4} \cline{5-6}
& & 140 K & 25 K  & GGA+SOC &GGA+U+SOC
\\ 
\hline \hline 
$A$-site &\ce{Co_A} &  ($0, 0, 2.18(6)$) & ($0, 0, 3.07(6)$) &($0, 0, 2.48$)  &($0.80, 0, 2.74$) \\ \hline 
&\ce{Co_B} &  NA & NA &($0, 0, -0.43$)  & ($-0.81, |0.13|, -2.72$) \\ 
$B$-site &\ce{V_B}  &  NA & NA &($0, 0, -1.29$) &($-0.34, |0.02|, -1.02$)\\ \cline{2-6}
&\ce{Co_B} + \ce{V_B} &    ($0, 0, -1.68(2)$) & ($|$0.88(8)$|$, $|$0.88(8)$|$, $-3.13(1)$) &($0, 0, -1.72$)  & ($-1.15, |0.15|, -3.74$) \\  \hline
 X-site& O & NA & NA &($0, 0, 0.09$)  & ($0, 0, 0.02$)\\
\hline
$\mu_{\rm {total}}\,(\SI{}{\micro_{\rm{B}}}/f.u.)$&  &   ($0, 0, 0.50(8)$) & ($0, 0, -0.06(7)$) & (0, 0, 1.14) &($-0.32, 0, -0.90$) \\
\end{tabular}
\end{ruledtabular}
\label{tab:Magn}
\end{table*}

\section{Conclusions}

Our interest in this less studied inverse cobalt vanadium spinel springs from the highly unconventional temperature-dependent magnetization of the compound, where a spontaneous magnetization reversal takes place. We have performed magnetization measurements and revealed at least three distinct anomalies. Neutron powder diffraction analysis reveals two antiparallel magnetic structures, each belonging to magnetic ions on $A$- and $B$-sites. The relative balance of these two structural components in essence determines the magnetization. However, the order parameter of the magnetic order on one site develops more moderately than that on the other site. While in itself unusual, a tipping of the magnetic balance, caused by such site-specific magnetic fluctuation, gives rise to a spontaneous flipping of the magnetization as the temperature is lowered. The magnetization reversal determined at zero fields is sensitive to an applied magnetic field, such that above $B\sim 0.25$\,T instead of a reversal a minimum in the magnetization is apparent at $T_{\rm MR}$.

The root cause of this site-specificity is unveiled by DFT results. The DFT calculations were performed assuming delocalized and localized spin carriers. We argue that the magnetization reversal is a consequence of the itineracy of Co and V electrons in the $B$-site. In particular, the {\it delocalized} calculations, which we associate with the behavior of the high temperature of the system (around $T_{\rm C}$), predict relatively small net magnetic moments on the  $B$-site compared to the $A$-site giving rise to the net ferrimagnetic moment observed near $T_{\rm C}$. The DFT also shows that the Co moments on the $A$-site are less sensitive to temperature change with electronic configuration Co$^{2+}(3d^7)$ with $S=3/2$. For the $B$-site, the DFT finds mixed oxidation states for Co and V  that at low temperatures tend towards their localized values, namely,  Co$^{2+}$ ($S=3/2) $ (as in the $A$-site)  and V$^{4+}$ ($S=1/2$).  This suggests that the  MR is likely driven by {\it delocalized} to {\it localized}  cross-over of the $3d$-electrons of the Co/V atoms in the $B$-site. We note that quantitative disagreements between theory and experiment are expected owing to the complexities of this inverse spinel Co$_2$VO$_4$ and the random distribution of Co and V in the $B$-site. Polarized muon spectroscopy as a function of temperature is consistent with the neutron diffraction and magnetization results but adds new insights that have been predicted by our DFT calculations. The muon spectroscopy indicates highly delocalized spin-carriers at high temperatures that gradually tend to localization at lower temperatures.

\section{Acknowledgement}

Part of this research at Ames Laboratory is supported by the U.S. Department of Energy, Office of Basic Energy Sciences, Division of Materials Sciences and Engineering under Contract No. DE-AC02-07CH11358. The electronic structure and magnetism  employed in this work are developed by D.P and his group in the Critical Materials Institute, an Energy Innovation Hub led by the Ames Laboratory and funded by the U. S. Department of Energy, Office of Energy Efficiency and Renewable Energy, Advanced Manufacturing Office. Use of the Spallation Neutron Source at the Oak Ridge National Laboratory is supported by the U.S. Department of Energy, Office of Basic Energy Sciences, Scientific Users Facilities Division. Experiments at the ISIS Pulsed Neutron and Muon Source were supported by a beamtime allocation from the Science and Technology Facilities Council. We thank the Helmholtz-Zentrum Berlin f\"ur Materialien und Energie for the allocation of neutron beamtimes at BER II and bulk properties measurements at  CoreLab Quantum Materials.

\bibliography{biblio}

\end{document}